# Insight into Potential Well Based Nanoscale FDSOI MOSFET Using Doped Silicon Tubs- A Simulation and Device Physics Based Study: Part I: Theory and Methodology

Shruti Mehrotra, S. Qureshi

*Department of Electrical Engineering, Indian Institute of Technology Kanpur, India*

**Abstract**

A novel planar device having doped silicon regions (tubs) under the source and drain of an FDSOI MOSFET is reported at 20 nm gate length. The doped silicon regions result in formation of potential wells (PW) in the source and drain regions of FDSOI MOSFET and thus, the device being called as Potential Well Based FDSOI MOSFET (PWFDSOI MOSFET). Simulation and device physics study on PWFDSOI MOSFET showed reduction in the OFF current of the device by orders of magnitude. A low $I_{OFF}$ of 22 pA/$\mu$m, high $I_{ON}/I_{OFF}$ ratio of 1.5 x $10^7$ and subthreshold swing of 76 mV/decade were achieved in 20 nm gate length PWFDSOI MOSFET. The study was performed on devices with unstrained silicon channel.

## 1. Introduction

The recent past has witnessed aggressive scaling of the technology node in a bid to meet Moore's law. This scaling has caused reduced gate control which consequently leads to high OFF current ($I_{OFF}$) and poor subthreshold swing (SS) [1]. Studies have reported devices like FinFET, TFET and the concept of negative capacitance to overcome these short comings [2, 3, 4, 5, 6, 7]. Even though FinFET offers excellent gate control, being a 3D device it has process and design level complexities. While NCFET improves SS with increasing thickness of the ferroelectric layer, it suffers from hysteresis [8]. The Fully Depleted Silicon-on-Insulator (FDSOI) MOSFET continues to be an attractive device option because of its planar topology and back-biasing feature. We have recently reported a novel Potential Well Based FDSOI MOSFET (PWFDSOI MOSFET) at 10 nm gate length to achieve a high ON-to-OFF current ratio ($I_{ON}/I_{OFF}$) [9]. In this paper, we explore the details of formation of potential wells at 20 nm gate length in the source and drain regions due to the introduction of heavily doped silicon regions in the BOX under the source and drain. The paper is based on simulation studies supported by p-n junction device physics responsible for the creation of potential wells in the source/drain regions which results in significant reduction in the OFF current of the device. This paper is organised as follows: Section 2 discusses the concept of formation of potential wells in the source and drain regions and the simulation methodology followed to realize the conventional FDSOI MOSFET with ground plane (GP)(reference device in this study) and the proposed PWFDSOI MOSFET at 20 nm gate length with doped silicon regions under the source and drain in the BOX. Section 3 discusses the physics behind PWFDSOI MOSFET and Section 4 draws the conclusion.

*Corresponding author

*Email address:* mshruti@iitk.ac.in (Shruti Mehrotra)

## 2. Formation of Potential Wells

We report a novel approach to create potential wells in the source and drain regions of PWFDSOI MOSFET by introducing doped silicon regions in the buried oxide under the source and drain regions. This study was performed on devices with unstrained silicon channel.

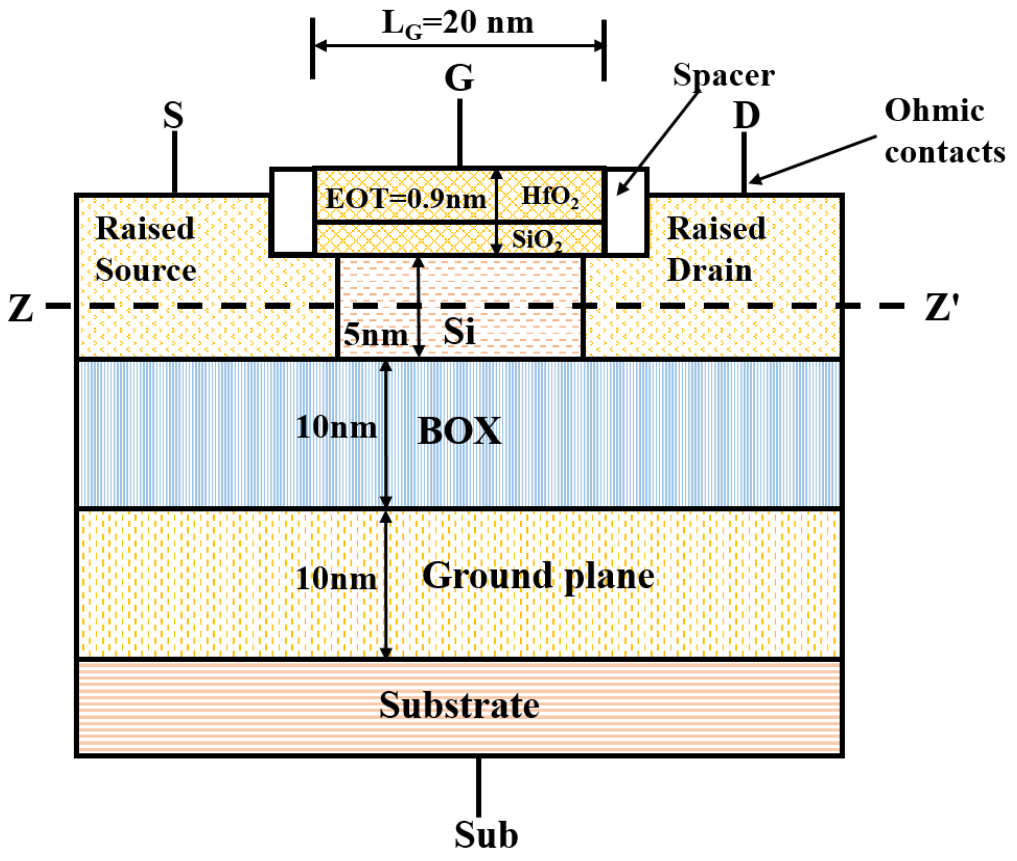

(a)

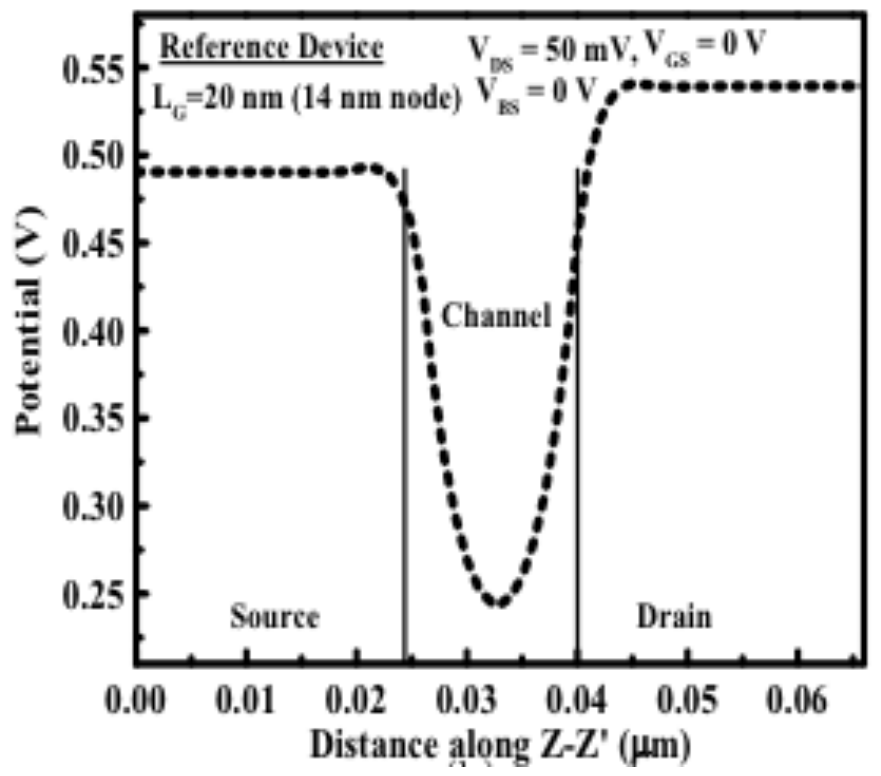

(b)

Fig. 1: (a) Schematic of 20 nm FDSOI MOSFET with $p^+$ ground plane (GP) under the BOX. The horizontal cutline Z-Z' is drawn in the center of the channel (2.5 nm from the gate stack/channel interface). (b) Relative channel potential profile in FDSOI MOSFET with GP in OFF state ($V_{GS}$= 0 V). Absence of potential wells in source and drain regions in 20 nm FDSOI MOSFET is observed.

Table 1: Models used in 2D TCAD simulations of FDSOI MOSFET

| **TCAD Models** | **Physical Effect Captured** |
|---|---|
| Drift Diffusion | Carrier transport |
| SRH and Auger | Carrier recombination |
| Quantum confinement | SOI layer is very thin leading to quantum confinement of carriers |
| Lombardi mobility model | Acoustic phonon scattering at low fields and surface recombination scattering at high transverse fields |
| High field mobility model | Velocity saturation effect |
| Self heating model | Lattice heating in the SOI layer |
| Fermi Dirac carrier statistics | Presence of heavily doped regions in the device |
| Bandgap narrowing | At very high doping in silicon, the $pn$ product becomes doping dependent |

Table 2: Device Parameters used in Simulations at 20 nm Gate Length

| **Parameter** | **Value** |
|---|---|
| Gate Length | 20 nm |
| EOT | 0.9 nm |
| $HfO_2$ thickness | 3.8 nm |
| $SiO_2$ thickness | 0.3 nm |
| Permittivity of $HfO_2$ | 25 |
| SOI layer thickness | 5 nm |
| BOX thickness | 10 nm |
| GP thickness | 10 nm |
| Spacer length | 3 nm |
| Gate-to-source/drain overlap | 3 nm |
| SOI layer doping | $10^{15}$ cm$^{-3}$ |
| Source/Drain doping | $10^{20}$ cm$^{-3}$ |
| $T_S/T_D$ doping | $10^{20}$ cm$^{-3}$ |
| Ground plane doping | $10^{20}$ cm$^{-3}$ |
| Substrate doping | $10^{15}$ cm$^{-3}$ |
| Work function of gate metal | 4.52 eV |

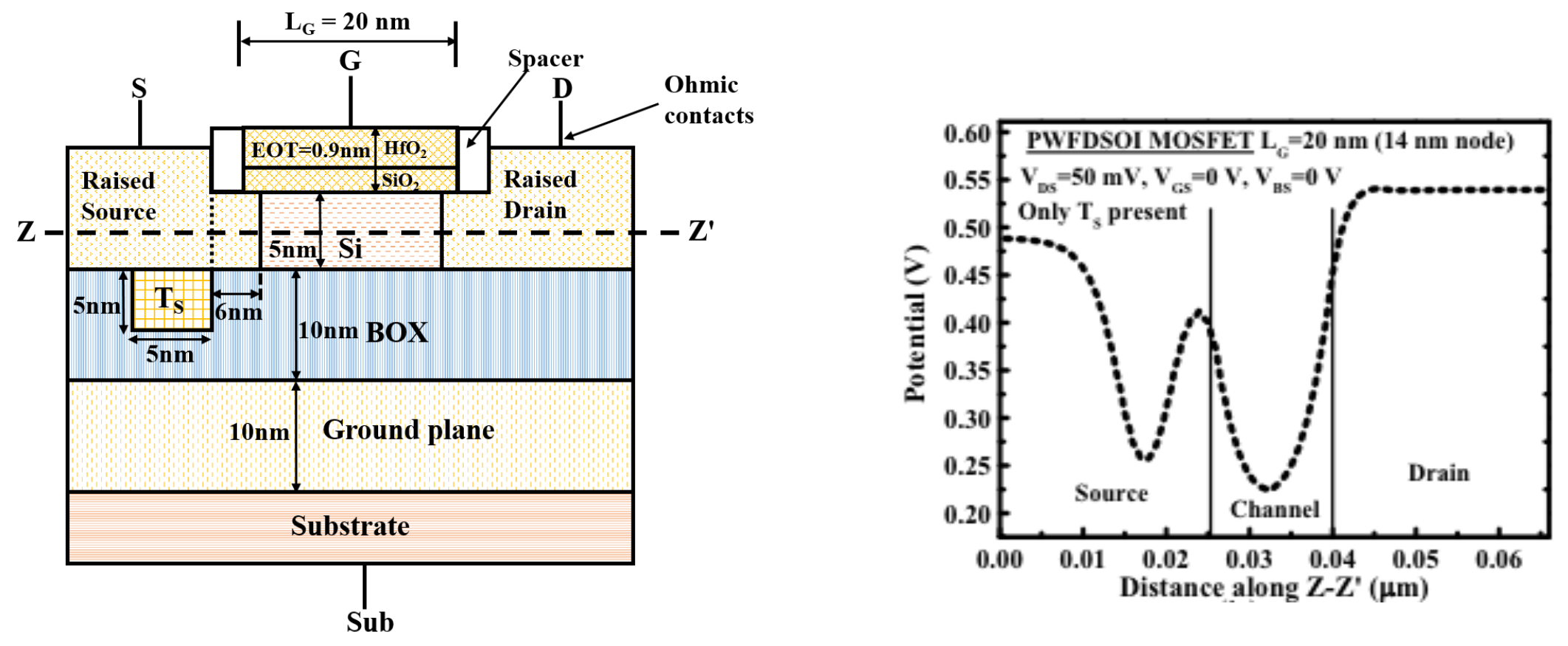

(a) (b)

Fig. 2: (a) Schematic of 20 nm PWFDSOI MOSFET with region $T_S$ under the source. The doping of $T_S$ is p-type with a concentration of 1 x $10^{20}$ $cm^{-3}$. The horizontal cutline Z-Z' is drawn in the center of the channel (2.5 nm from the gate stack/channel interface). (b) Potential well in the source region of 20 nm PWFDSOI MOSFET with region $T_S$ under the source. The relative potential profile is plotted along cutline Z-Z' when the device is in OFF state ($V_{GS}$ = 0 V).

### 2.1. *Simulation Methodology and Reference Device*

FDSOI MOSFET with degenerately doped (~$10^{20}$ $cm^{-3}$) region under the BOX called the ground plane (GP) was the reference device in this study [10]. First a structure as proposed in [10] was realized at 20 nm gate length for calibration. The silicon layer that forms the channel was 6 nm thick and the BOX was 25 nm thick. As a next step to realize an ultra thin body and BOX structure, the silicon layer thickness was reduced to 5 nm and the BOX thickness was reduced to 10 nm. The device had an HKMG gate stack with an EOT of 0.9 nm as shown in Fig. 1(a). The gate-to-source/drain overlap was 3 nm. This has also been confirmed through process simulations in Silvaco Athena [11]. As shown in Fig. 1(b) the relative potential profile along the cutline Z-Z' through the center of the channel shows the channel potential as expected. The simulation study was performed in Silvaco TCAD [12]. Relevant models invoked to capture the physics of the devices are mentioned in Table 1. The device parameters used in simulations are mentioned in Table 2.

### 2.2. *20 nm gate length PWFDSOI MOSFET*

The proposed PWFDSOI MOSFET had heavily doped ($10^{20}$ $cm^{-3}$) silicon regions (tubs) $T_S$ and $T_D$ under the source and drain respectively in the BOX. The proposed device PWFDSOI MOSFET was identical to the reference device in all respects except for the presence of the doped silicon tubs $T_S$ and $T_D$. The doping of the tubs was of opposite type as compared to source/drain doping ($p^+$ in n-channel PWFDSOI MOSFET). These doped silicon tubs were 5 nm (depth) x 5 nm (width). As mentioned earlier, the presence of $T_S$ and $T_D$ leads to p-n junctions formation and consequent potential wells in the source and drain regions.

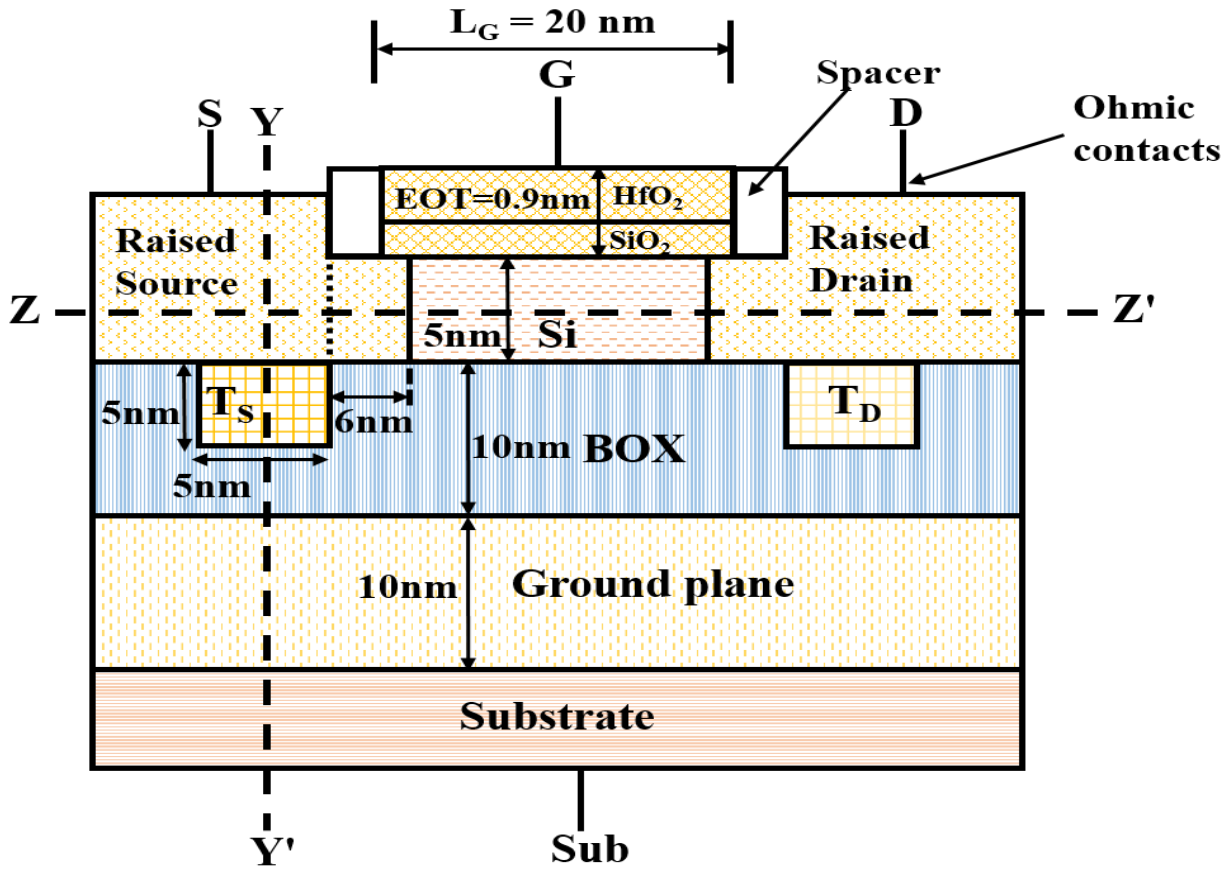

Fig. 3: 20 nm PWFDSOI MOSFET with regions $T_S$ and $T_D$ under the source and drain respectively. The doping of $T_S$ and $T_D$ is p-type with a concentration of 1 x $10^{20}$ $cm^{-3}$. It has an HKMG gate stack with an EOT of 0.9 nm. The cutline Z-Z' is through the center of the channel.

Figure 2(a) shows 20 nm PWFDSOI MOSFET when only $T_S$ is present under the source. Relative potential profile along cutline Z-Z' for this case shows the potential well in the source only which is illustrated in Fig. 2(b). Similarly, when only $T_D$ is present under the drain, a potential well is formed in the drain only.

The schematic of the proposed 20 nm PWFDSOI MOSFET is shown in Fig. 3. Figure 4(a) shows the potential wells in the source and drain regions for the proposed 20 nm gate length PWFDSOI MOS-

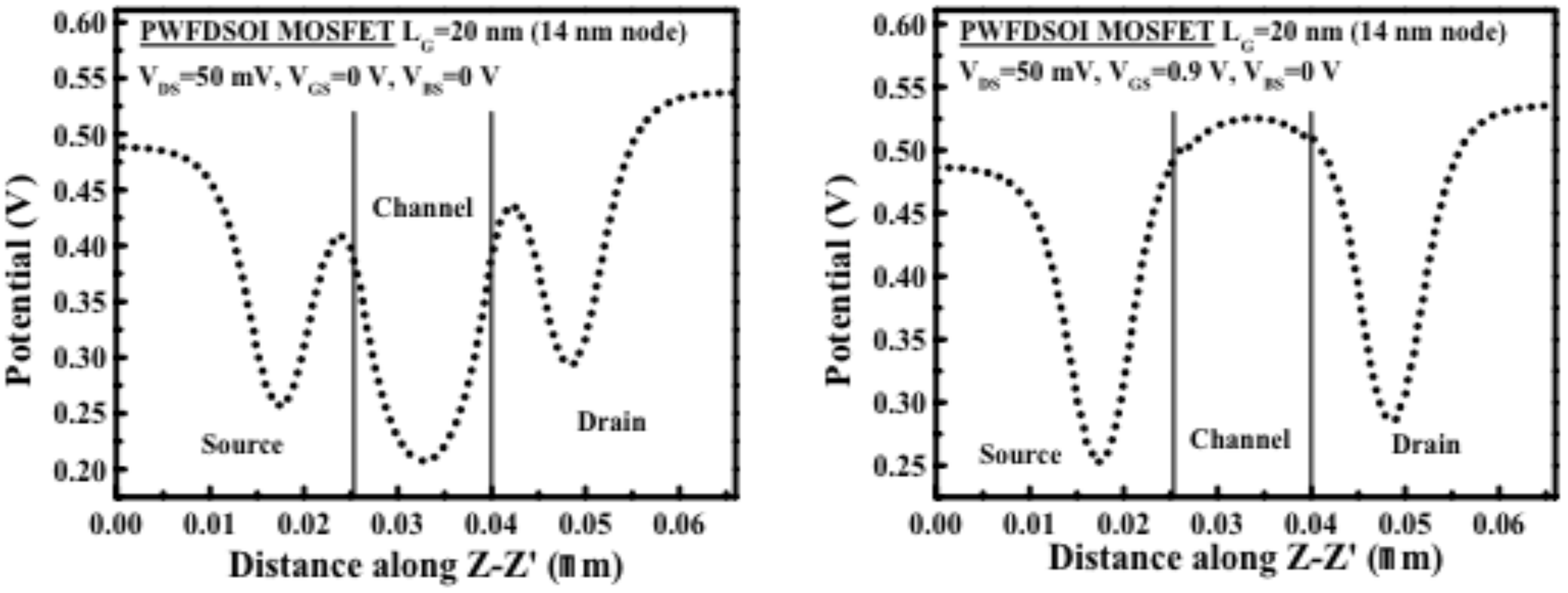

(a) (b)

Fig. 4: Potential wells in the source and drain regions of 20 nm PWFDSOI MOSFET with regions $T_S$ and $T_D$ under the source and drain respectively. The relative potential profile is plotted along cutline Z-Z' when the device is in (a) OFF state ($V_{GS}$= 0 V), and, (b) ON state ($V_{GS}$= 0.9 V). $V_{DS}$= 50 mV, $V_{BS}$= 0 V.

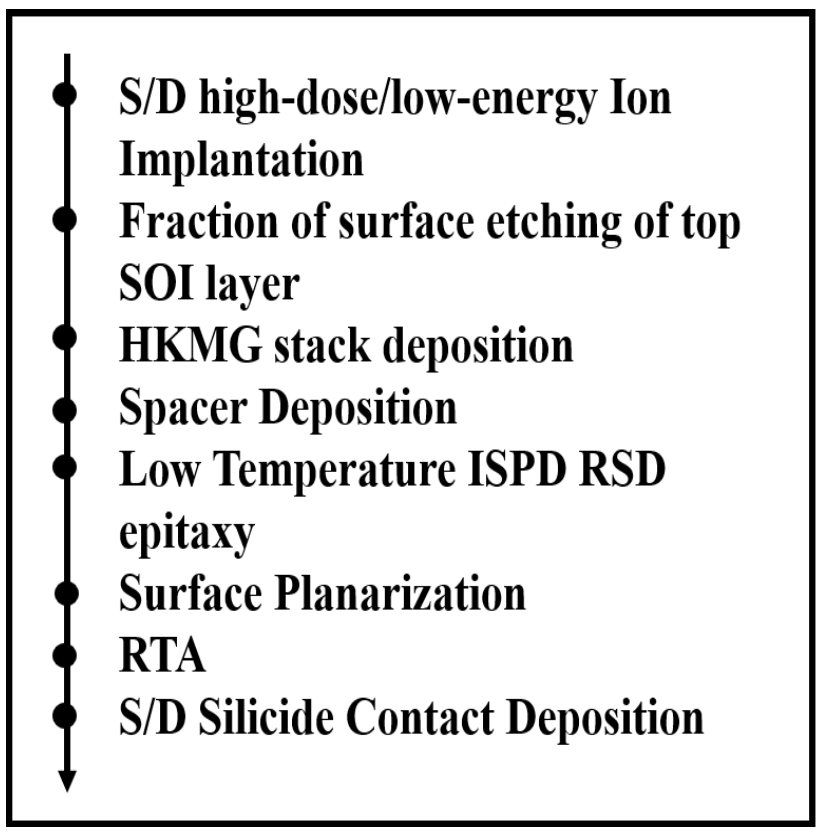

Fig. 5: Proposed process flow for the fabrication of PWFD-SOI MOSFET from reference device [11].

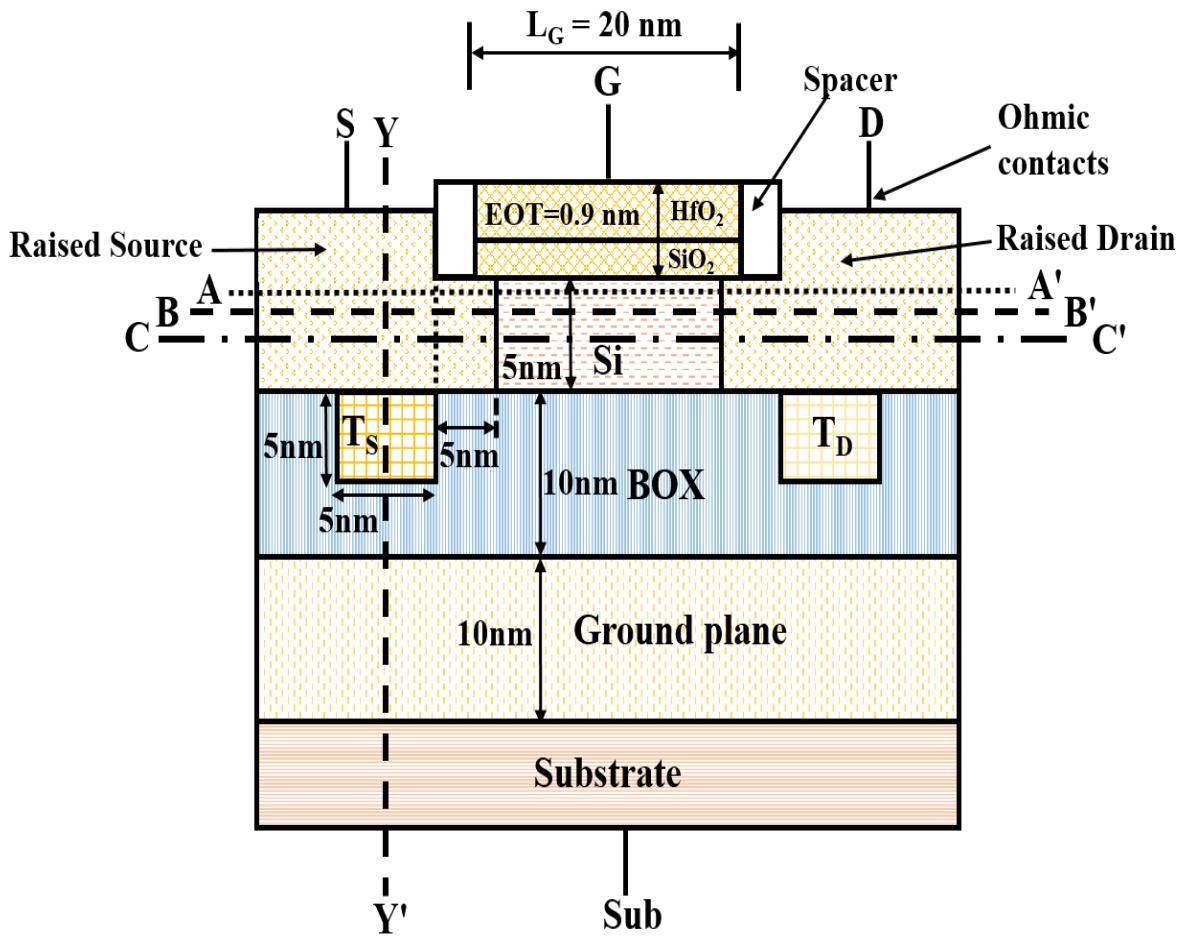

Fig. 6: Schematic of 20 nm gate length PWFDSOI MOSFET with BOX thickness of 10 nm. The cutline A-A' is taken below the gate stack/channel interface at a distance of 0.5 nm from the gate stack/channel interface, the cutline B-B' is taken at a distance of 1 nm from the gate stack/channel interface and the cutline C-C' is taken below the gate stack/channel interface at a distance of 1.5 nm from the gate stack/channel interface. The cutline Z-Z' is taken in the center of the SOI layer as shown in Fig. 3. The cutlines are not drawn to scale.

FET (Fig. 3) along the same cutline Z-Z' for OFF state of the device ($V_{DS}$ = 50 mV, $V_{GS}$ = 0 V and $V_{BS}$ = 0 V). Fig. 4(b) shows the relative potential profile along Z-Z' for ON state of the device ($V_{DS}$ = 50 mV, $V_{GS}$ = 0.9 V and $V_{BS}$ = 0 V). The channel is clearly seen to be in inversion. The proposed process flow for the PWFDSOI MOSFET is shown in Fig. 5. The details are mentioned in [11]. Simulation studies performed using Silvaco Athena and Atlas showed diffusion of dopants from tubs is controllable and no dopants diffuse into the channel [11].

### 2.3. *Validity of Drift Diffusion Transport Model*

For OFF state of the device, the relative potential profile was also studied along three more horizontal cutlines A-A', B-B' and C-C' apart from Z-Z' as shown in Fig. 6. The cutline A-A' is taken below the gate stack/channel interface at a distance of 0.5 nm from the interface, the cutline B-B' is taken at a distance of 1 nm from the gate stack/channel interface and the cutline C-C' is taken at a distance of 1.5 nm from the gate stack/channel interface. The relative potential profiles of PWFDSOI MOSFET along these cutlines are shown in Fig. 7. It can be seen from Fig. 7(a) that the potential well depth seen by electrons in the top portion of the source is small. The potential well depth in source along A-A' is 16 meV, along B-B' is 42 meV and along C-C' it is 86 meV. The potential well depth becomes significantly large at the center along the cutline Z-Z' having a value of 225 meV as observed from Fig. 4(a). Therefore, it is reasonable to assume that primarily electrons in the top portion of the source in the vicinity of the gate contribute to the OFF current in PWFDSOI MOSFET. Under this assumption, the carrier transport mechanism remains the same as in the reference device. Thus, the drift-diffusion transport model is used in the TCAD simulations of the PWFDSOI MOSFET also under this assumption because in PWFDSOI MOSFET it is the reduction in the number of carriers contributing to the OFF current as compared to the reference device.

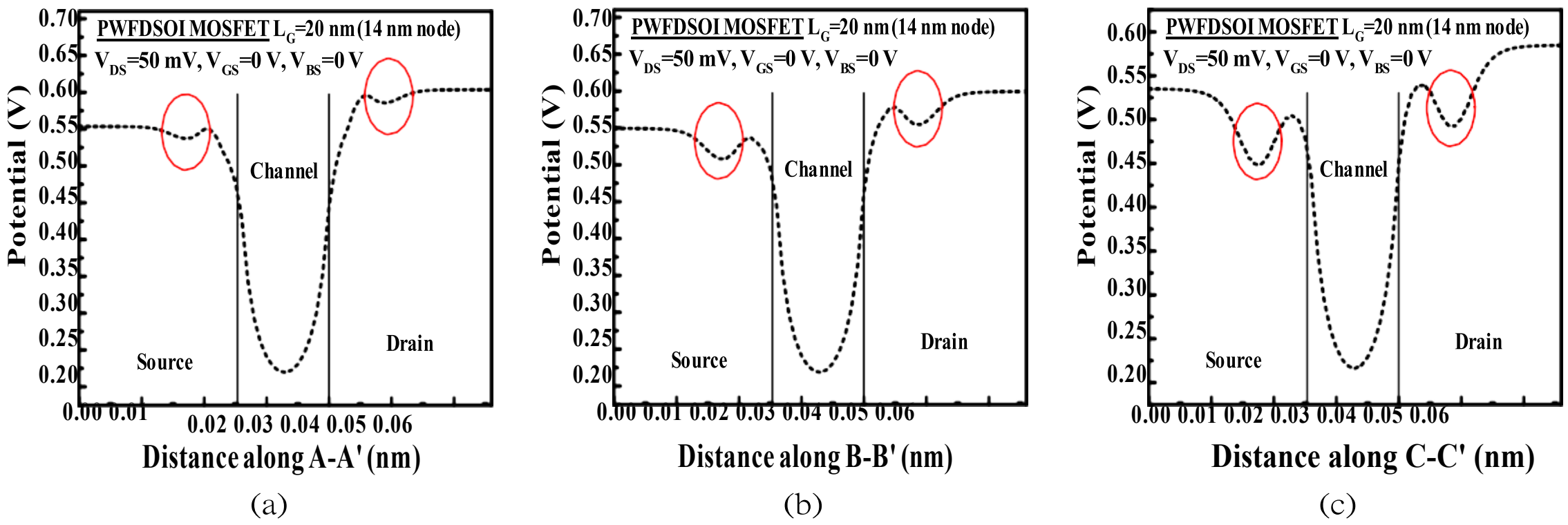

Fig. 7: (a)Relative potential profile along cutline A-A' in OFF state for 20 nm gate length PWFDSOI MOSFET. The potential well depth in the source is very small (16 meV). (b) Relative potential profile along cutline B-B' in OFF state for 20 nm gate length PWFDSOI MOSFET. The potential well depth in the source is 42 meV. (c) Relative potential profile along cutline C-C' in OFF state for 20 nm gate length PWFDSOI MOSFET. The potential well depth in the source is 86 meV. The circled portions show the potential wells in the source and drain regions. $V_{DS}$ = 50 mV, $V_{GS}$ = 0 V and $V_{BS}$ = 0 V.

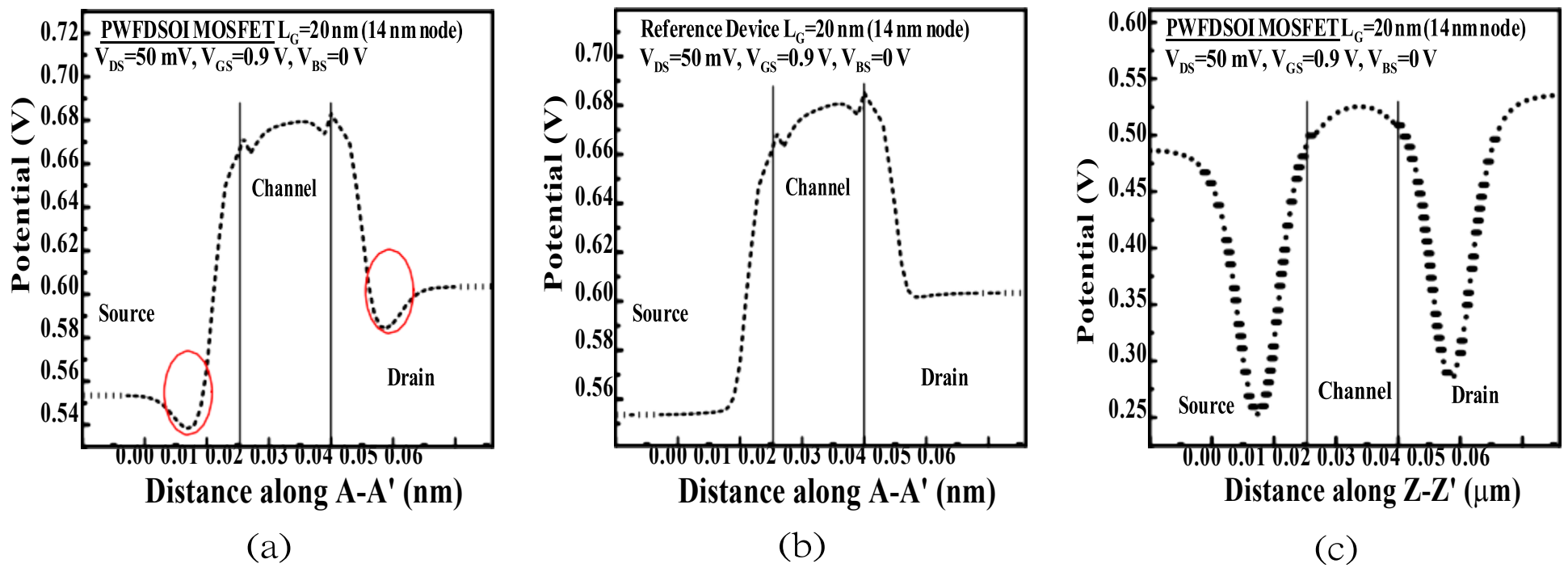

Fig. 8: (a)Relative potential profile along cutline A-A' in ON state for PWFDSOI MOSFET. The potential well depth in source and drain regions is very small. (b) Relative potential profile along cutline A-A' in ON state for reference FDSOI MOSFET. There is absence of potential wells in source and drain regions in the reference device. (c) Potential wells in source and drain regions along cutline Z-Z' in 20 nm PWFDSOI MOSFET when device is in inversion ON state. $V_{DS}$ = 50 mV, $V_{GS}$ = 0.9 V and $V_{BS}$ = 0 V.

For PWFDSOI MOSFET in the ON state ($V_{DS}$ = 50 mV, $V_{GS}$ = 0.9 V, $V_{BS}$ = 0 V) of the device, the relative potential profile has been analysed along two cutlines, A-A' and Z-Z' as shown in Figs. 6 and 3 respectively. The relative potential profile along cutline A-A' for PWFDSOI MOSFET in ON state is shown in Fig. 8(a). The potential well depth in the source and drain regions is very small, same as in OFF state (16 meV). The relative potential profile along the same cutline A-A' for the reference device is shown in Fig. 8(b). The relative potential profile along cutline Z-Z' in 20 nm gate length PWFDSOI MOSFET in ON state of the device is shown in Fig. 8(c). The potential well depth in the source and drain regions is more than that observed in Fig. 8(a) and is around 300 meV.

## 3. Physics of PWFDSOI MOSFET

### 3.1. *Transfer characteristics*

#### 3.1.1. *20 nm gate length PWFDSOI MOSFET*

The transfer characteristics of the 20 nm gate length PWFDSOI MOSFET are shown in Fig. 9 at $V_{DS}$ = 0.9 V and in the absence of a back-bias. It is interesting to note that $I_{OFF}$ ($I_D$ at $V_{GS}$ = 0 V) when only $T_D$ was present was less than when only $T_S$ was present. This is because when only $T_S$ was present, the carriers which have sufficient energy to overcome the source-to-channel barrier reach the drain terminal and contribute to $I_{OFF}$. On the contrary, when only $T_D$ was present smaller $I_{OFF}$ resulted as the carriers experienced barrier due to $T_D$ even though more carriers from the source were expected to get into the channel. Consequently, when neither $T_S$ nor $T_D$ was present (Figs. 1(a) and 1(b)), $I_{OFF}$ was maximum and when both $T_S$ and $T_D$ were present (Figs. 3 and 4(a)), $I_{OFF}$ was minimum. The presence of potential wells in source and drain does not lead to a significant reduction in ON current ($I_{ON}$ = $I_D$ at $V_{GS}$ = $V_{DS}$ ) as compared to ON current of the reference device. This results in improved $I_{ON}/I_{OFF}$ ratio for 20 nm PWFDSOI MOSFET.

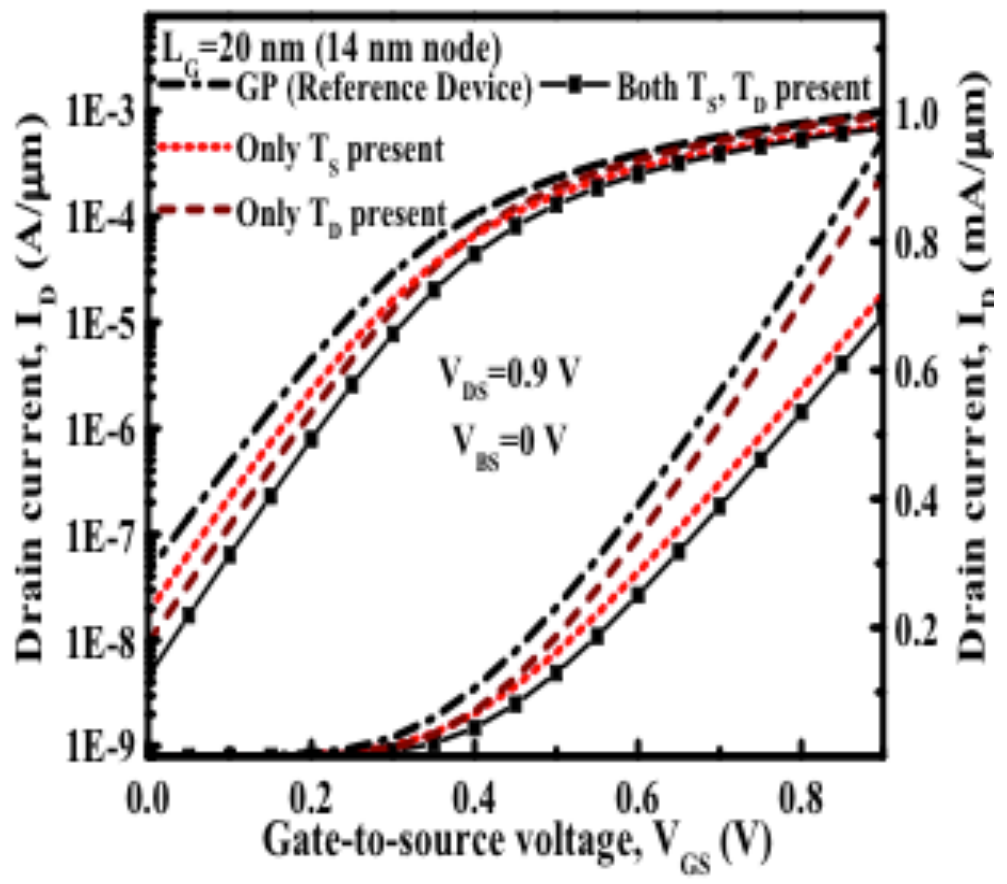

Fig. 9: $I_D$ vs. $V_{GS}$ characteristics of 20 nm PWFDSOI MOSFET with regions $T_S$ and $T_D$ under the source and drain respectively. A step-by-step comparison is drawn in the figure between reference device, either $T_S$ or $T_D$ present and the case when both $T_S$ and $T_D$ are present (proposed device). $V_{DS}$ = 0.9 V and $V_{BS}$ = 0 V.

Further, a back-bias of -1 V was applied to achieve better front gate control [13]. In PWFDSOI MOSFET, besides improving gate control, back-bias was seen to increase the depth of the potential wells and thus, reduction of OFF current. Figure 10 shows the transfer characteristics of the 20 nm PWFDSOI MOSFET and the reference device in the presence of back-bias. Table 3 gives the device performance parameters for 20 nm PWFDSOI MOSFET for the bias condition $V_{DS}$ of 0.9 V and $V_{BS}$ of -1.0 V.

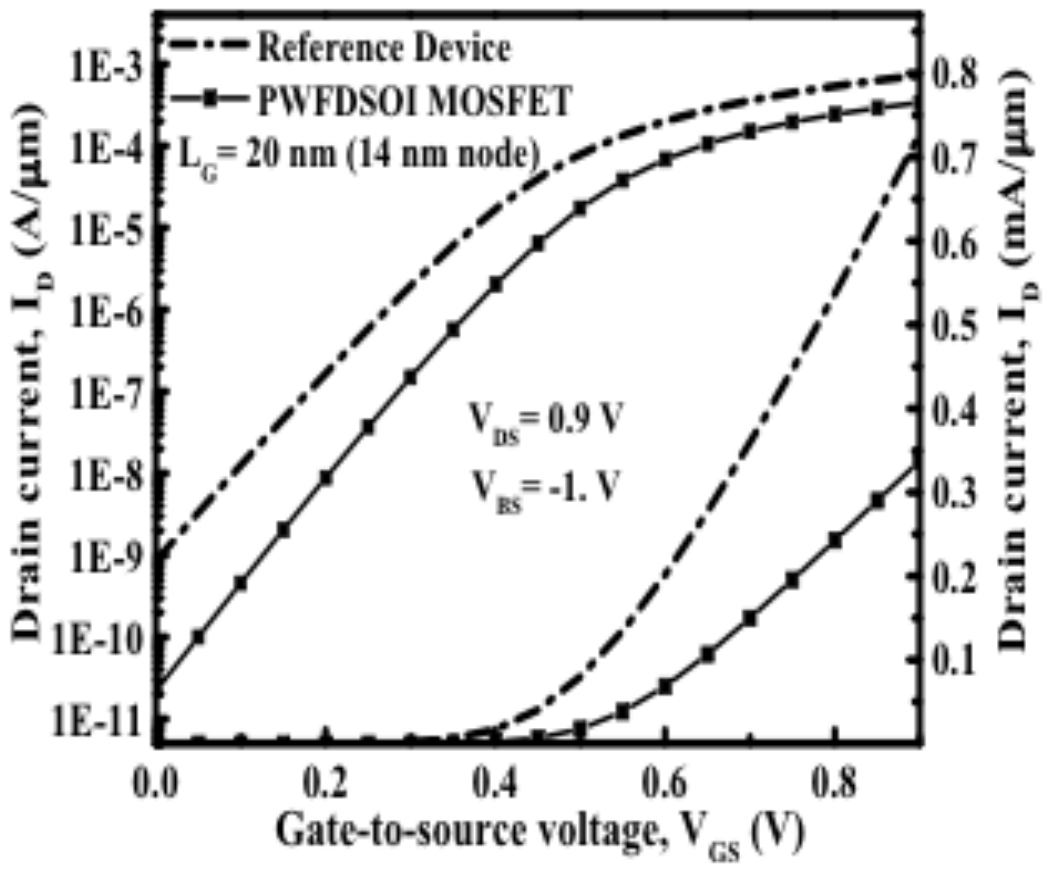

Fig. 10: $I_D$ vs. $V_{GS}$ characteristics of 20 nm PWFDSOI MOSFET with regions $T_S$ and $T_D$ under the source and drain respectively. A comparison is drawn with the reference device. $V_{DS}$ = 0.9 V and $V_{BS}$ = -1.0 V.

#### *3.1.2. Physics of PWFDSOI MOSFET*

### *3.2. Electric Field*

The electric field profile in PWFDSOI MOSFET is altered significantly in comparison to the electric field profile of the reference device as shown in Fig. 11. This can be attributed to the presence of space charge created by the formation of the p-n junctions which makes the presence of the GP more effective in the termination of field lines. This also explains the improvement of DIBL in PWFDSOI MOSFET in comparison to GP as discussed in Part II.

### *3.3. Potential well depth as function of distance along Y-Y′ in the source*

The variation of depth of potential well in the source as we move from source towards source/$T_S$ interface is shown in Fig. 12. The potential well depth increases as we move along cutline Y-Y' from a point in the source to the source/$T_S$ interface. This can be attributed to the increased influence of the positive space charge on the carriers. Thus, electrons located deeper in the source are less likely to contribute to the OFF current. With the application of a back-bias, the potential well depth increases further causing a significant reduction in leakage current as shown in Fig. 10 and for reasons mentioned in Part II.

### *3.4. Potential variation along Y-Y′*

The variation of relative potential was also studied along the cutline Y-Y' across the n-p junction formed by the source and $T_S$ in 20 nm PWFDSOI MOSFET and is shown in Fig. 13. The study was performed under equilibrium condition

Table 3: Performance parameters at 20 nm gate length

| | **PWFDSOI MOSFET** | **Reference Device** |
|---|---|---|
| SS (mV/decade) | 76 | 85 |
| $I_{OFF}$ (pA/$\mu$m) | 22 | 860 |
| $I_{ON}$ (mA/$\mu$m) | 0.34 | 0.37 |
| $I_{ON}/I_{OFF}$ | $1.5 \times 10^7$ | $8.4 \times 10^5$ |

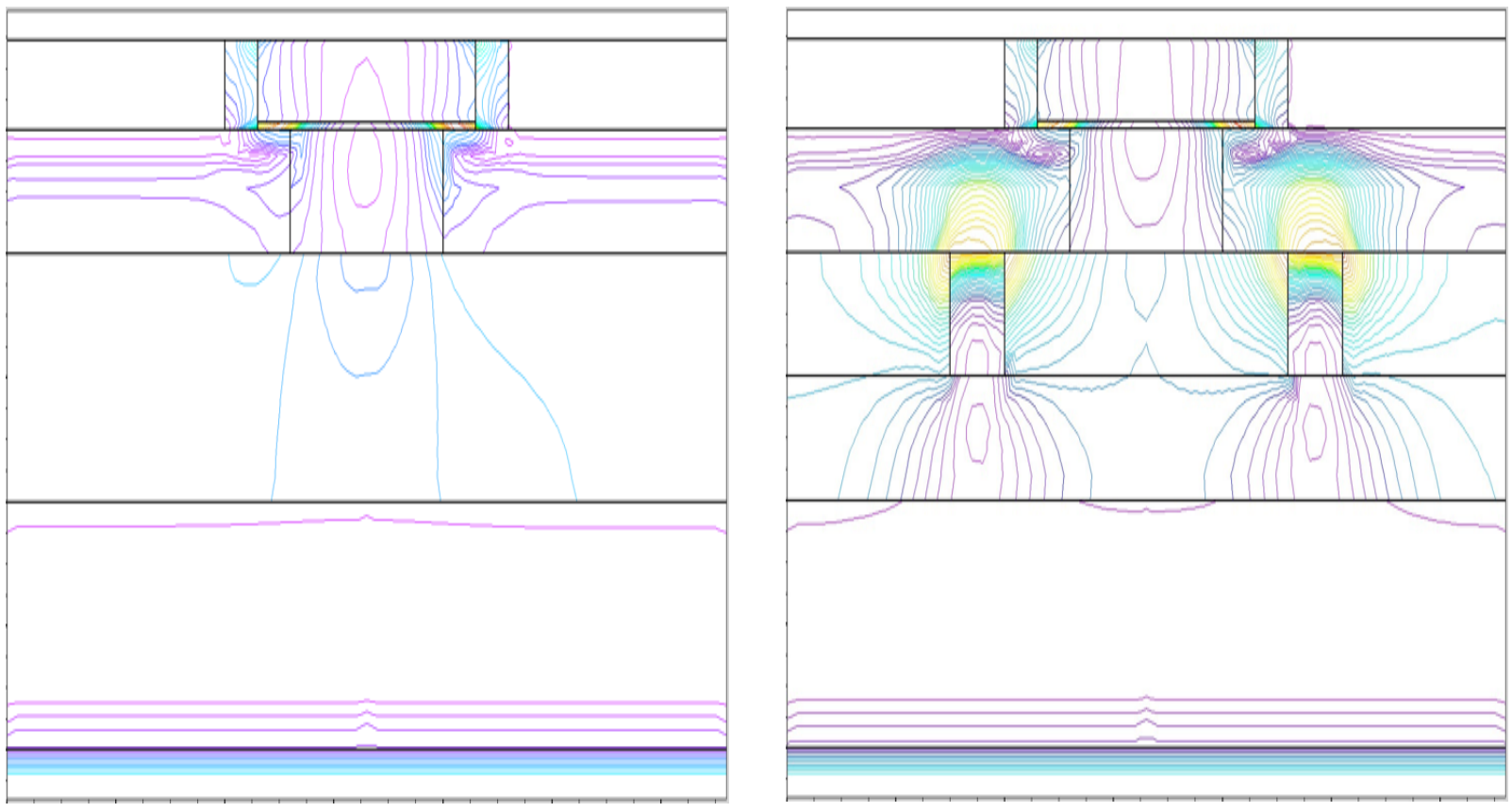

(a) (b)

Fig. 11: Electric field profiles of (a) reference device and, (b) PWFDSOI MOSFET. $V_{DS}$ = 50 mV, $V_{GS}$ = 0 V and $V_{BS}$ = 0 V.

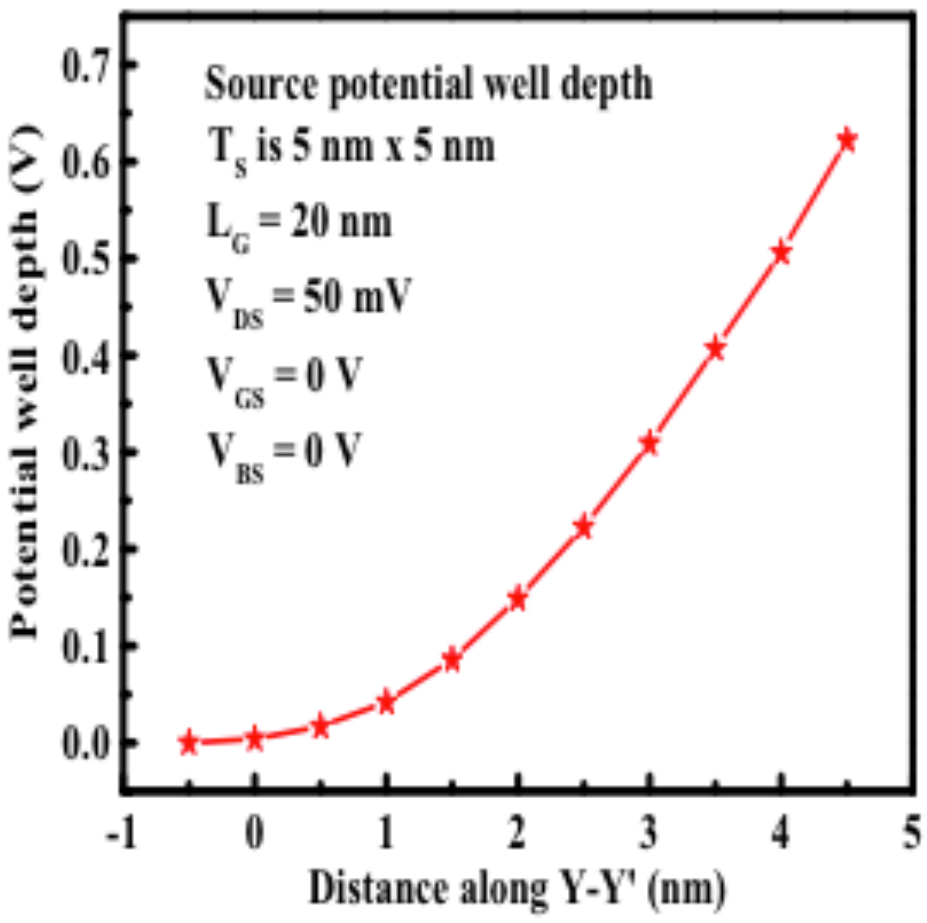

Fig. 12: Variation of potential well depth in the source in 20 nm PWFDSOI MOSFET along cutline Y-Y' . Here Y-Y' =0 is the gate stack/channel interface as shown in Fig. 3.

($V_{DS}$ =$V_{GS}$ =$V_{BS}$ =0 V). The difference in the potential across the n-p junction is approximately equal to 1.12 eV which is the band gap energy of silicon as shown in Fig. 13. Also, it is clearly observed in Fig. 13 that a significantly larger potential drop occurs in the source region and only about 2 nm depth of $T_S$ from the source/$T_S$ interface is depleted. This suggests significant depth of $T_S$ region is quasi neutral. This implies that process variations in depth of the $T_S$ and $T_D$ will not make a significant impact on the performance of the device.

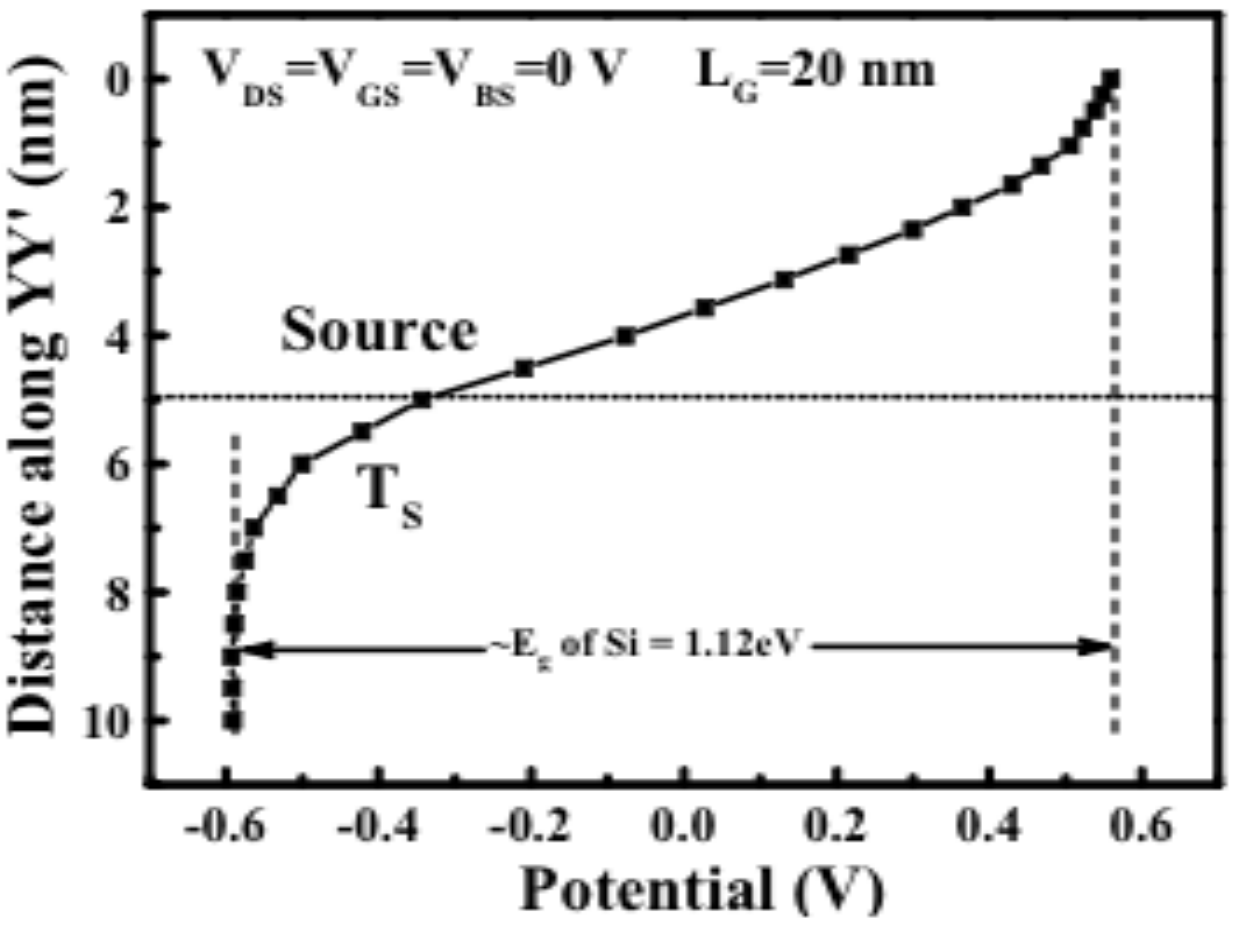

Fig. 13: Variation of relative potential from source to $T_S$ in 20 nm PWFDSOI MOSFET under the condition: $V_{DS}$ = $V_{GS}$ = $V_{BS}$ = 0 V for doped silicon $T_S$ and $T_D$. Here Y-Y' =0 is the gate stack/channel interface as shown in Fig. 3.

## 4. Conclusion

A study based on simulation and device physics concepts of 20 nm gate length PWFDSOI MOSFET has been presented in this paper. The potential wells are instrumental in reducing the OFF current of the device significantly. However, the reduction in ON current is marginal, thus significantly improving $I_{ON}/I_{OFF}$ ratio. The study was done on devices with unstrained silicon channel. However, the study is also applicable to scaled devices having strained silicon channel, thus expecting improvement in $I_{ON}$ and $I_{ON}/I_{OFF}$ ratio.